\documentclass[twocolumn,showpacs,preprintnumbers,amsmath,amssymb]{revtex4}

\usepackage{graphicx}
\usepackage{dcolumn}
\usepackage{bm}

\begin{document}

\preprint{APS/123-QED}

\title{Origin of coexisting large Seebeck coefficient 
and metallic conductivity in the electron doped SrTiO$_3$ and KTaO$_3$
}

\author{Hidetomo Usui}
\author{Shinsuke Shibata\cite{Shibata}}%
\author{Kazuhiko Kuroki}%
\affiliation{Department of Applied Physics and Chemistry,
The University of Electro-Communication, Chofu, Tokyo 182-8585, Japan}%

\date{\today}

\begin{abstract}
We study the origin of the large Seebeck coefficient despite the 
metallic conductivity in the La-doped SrTiO$_3$ and Ba-doped KTaO$_3$.
We calculate the band structure of SrTiO$_3$ and KTaO$_3$, 
from which the Seebeck coefficient is obtained using the Boltzmann's equation.
We conclude that the multiplicity of the $t_{2g}$ bands in these materials  
is one major origin of the good thermoelectric property 
in that when compared at a fixed total number of doped electrons, 
the Seebeck coefficient and thus the power factor are larger in 
multiple band systems than in single band ones because the 
number of doped electron bands {\it per band} is smaller in the former. 
We also find that the second nearest neighbor hopping integral,
which generally has  negative values in these materials and works 
destructively against the Seebeck effect, is nearly 
similar between KTaO$_3$ and SrTiO$_3$ despite the larger band width in 
the former. This can be another factor favorable for thermopower in 
the Ba-doped KTaO$_3$.

\end{abstract}

\maketitle

\section{Introduction}

The discovery of the large Seebeck coefficient in
Na$_x$CoO$_2$\cite{Terasaki} and the findings in
cobaltates/cobaltites\cite{Li,Fujita,Hebert,Miyazaki,Lee} and
rhodates\cite{Okada,Okamoto} that followed have brought up an
interesting possibility of finding good thermoelectric
materials that have relatively high (metallic) conductivity.
These cobaltates and rhodates are materials where holes 
are doped into the $d^6$ configuration, namely the electron 
configuation where the $t_{2g}$ bands are fully filled.
On the other hand, there is another class of $t_{2g}$ transition 
metal oxides where relatively good thermoelectric properties are obtained, 
namely the electron doped materials such as SrTiO$_3$\cite{Okuda_SrTiO3}.
When Sr is partially replaced by La in SrTiO$_3$,  electrons are 
doped in the originally $d^0$ configuration.
This material exhibits 
large Seebeck coefficient despite showing metallic conductivity, 
and the power factor, i.e., the Seebeck coefficient squared times the 
conductivity, is comparable to that of Bi$_2$Te$_3$.
Quite recently, good thermoelectric properties have also been 
observed in Ba-doped KTaO$_3$\cite{Sakai_KTaO3}. This is 
another $t_{2g}$ oxide, where electrons are doped into the 
originally $d^0$ configuration, but is a 
$5d$ system as compared to the $3d$ 
in SrTiO$_3$. Here again, 
relatively large Seebeck coefficient is observed despite the metallic 
conductivity. 

Theoretically, there have been several approaches that 
explain the large Seebeck coefficient in oxides.
From the first principles band calculation studies, 
it has been pointed out that the narrowness of 
the band structure in Na$_x$CoO$_2$ and related rhodates 
is an important factor\cite{Singh,Wilson}.
We have proposed that besides the width of the band, 
the shape of the band, which we call the ``pudding-mold'' type band, 
is important for the coexistence of the large Seebeck coefficient
and the high conductivity in Na$_x$CoO$_2$\cite{Kuroki} and 
related rhodates\cite{Usui,AritaLiRh}.
On the other hand, Koshibae {\it et al.}  derived a formula for the  
Seebeck coefficient in the $T(\rm temperature)\rightarrow\infty$ limit, 
and pointed out that 
the orbital degeneracy originates large entropy, leading to the 
large Seebeck coefficient\cite{Koshibae,KoshibaePRL}.

In the present study, we propose that yet another mechanism, 
where the band multiplicity plays an important role, 
is at work in the {\it electron doped} $t_{2g}$ materials.
Namely, when there are multiple (nearly) equivalent 
bands at the Fermi level, and the number of doped 
electrons {\it per band} is fixed, 
the Seebeck coefficient is essentially the same 
regardless of the number of bands,
while the conductivity increases with the multiplicity of the 
bands thus resulting in an enhanced power factor. In other words, 
when the {\it total number of doped electrons} itself is 
fixed, the Seebeck coefficient and thus the power factor is larger for 
multiple band systems because the Fermi energy stays low. 
We also examine the 
effect of the band shape, and show that the second nearest 
neighbor hopping integral,
which generally has  negative values in these materials and work 
destructively against the Seebeck effect, is nearly 
similar between KTaO$_3$ and SrTiO$_3$ despite the larger band width in 
the former. This can be another factor favorable for good 
thermoelectric properties in the Ba-doped KTaO$_3$.

%

\section{Formulation}

\subsection{Boltzmann's equation approach}
We first briefly summarize the Boltzmann's equation approach 
adopted in the present study\cite{Mahan,Wilson}.
In this approach, the Seebeck coefficient is given as
\begin{equation}
{\bf S}=\frac{1}{eT}{\bf K}_0^{-1}{\bf K}_1
\label{eq1}
\end{equation}
where $e(<0)$ is the electron charge, $T$ is the temperature,
tensors ${\bf K}_0$ and ${\bf K}_1$ are given by
\begin{equation}
{\bf K}_n=\sum_{\Vec{k}}\tau(\Vec{k})\Vec{v}(\Vec{k})\Vec{v}(\Vec{k})
\left[-\frac{\partial f(\varepsilon)}
{\partial \varepsilon}(\Vec{k})\right]
(\varepsilon(\Vec{k})-\mu)^n.
\label{eq2}
\end{equation}
Here, $\varepsilon(\Vec{k})$ is the band dispersion, 
$\Vec{v}(\Vec{k})=\nabla_{\Vec{k}}\varepsilon(\Vec{k})$ is the
group velocity, $\tau(\Vec{k})$ is the quasiparticle lifetime,  
$f(\varepsilon)$ is the Fermi distribution function,  
and $\mu$ is the chemical potential. 
Hereafter, we simply refer to $({\bf K}_n)_{xx}$ as $K_n$, and
$S_{xx}=(1/eT)\dot(K_1/K_0)$ (for diagonal ${\bf K}_0$) as $S$.
Using $K_0$, conductivity can be given as
$\sigma_{xx}=e^2K_0\equiv\sigma={1/\rho}$.
As an input of the band structure in this calculation, 
we use the first principles calculation as described below.
$\tau(\Vec{k})$ will be 
taken as an (undetermined) 
constant in the present study, so that it cancels out 
in the Seebeck coefficient, while the conductivity and thus the 
power factor has to be normalized by a certain reference.

\subsection{Band Calculation}
SrTiO$_3$ and KTaO$_3$ have cubic perovskite structures.
We use the  experimentally determined lattice
constants in the band calculation, which are 
$a = 3.90528$\AA \ for SrTiO$_3$\cite{Mitchell_SrTiO3_a} 
and $a = 3.9883$\AA \ for KTaO$_3$\cite{Zhurova_KTaO3_a}.
For SrTiO$_3$, we have obtained the band structure using the
Quantum-ESPRESSO package\cite{pwscf}. 
In order to obtain a tight binding model on which we can perform various 
analysis, we construct  
the maximally localized Wannier functions (MLWFs)\cite{MaxLoc}
for the outer energy window $0{\rm eV} < \epsilon_{k}-E_F < 5$eV and for
the inner windows $0{\rm eV} < \epsilon_{k}-E_F < 4$eV,
where $\epsilon_{k}$ is the eigenenergy of the Bloch states
and $E_{F}$ is the Fermi energy. These MLWFs, centered at Ti sites
in the unit cell, have three orbital symmetries (orbital
1:$d_{xy}$, 2:$d_{yz}$, 3:$d_{zx}$). With these effective
hoppings and on-site energies, the tight-binding Hamiltonian
is obtained, and finally the Seebeck coefficient
is calculated using eq.(\ref{eq1}).
%
%
For KTaO$_3$, we have obtained the band structure 
using the WIEN2K package\cite{wien2k}.
The Seebeck coefficient is calculated using the BoltzTraP code\cite{boltztrap}.

\section{Calculation Results of the Seebeck coefficient}
In this section, we present the band calculation results and the 
calculation results of the Seebeck coefficient.
\begin{figure}
\includegraphics[width=8.5cm]{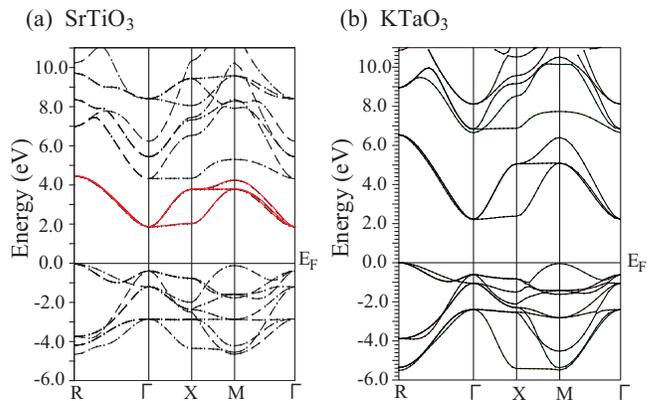}
\caption{\label{fig:1} The band structure of (a) SrTiO$_3$ 
and (b) KTaO$_3$ In (a), the black dotted lines are the original 
LDA calculation while the solid red lines are the bands of the 
tight binding model obtained using the MLWFs.}
\end{figure}
\begin{figure}
\includegraphics[width=8.5cm]{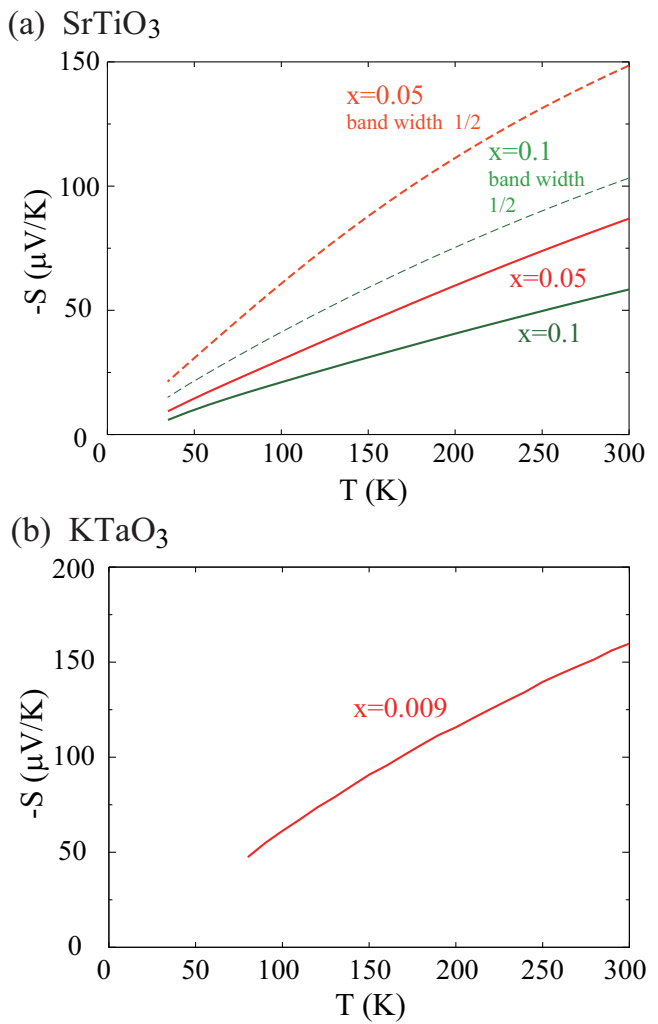}
\caption{\label{fig:2} The calculated 
Seebeck coefficient for (a) SrTiO$_3$ and (b) KTaO$_3$ plotted 
as functions of temperature for various doping rate $x$. }
\end{figure}
The calculated band structures of SrTiO$_3$ and KTaO$_3$ are shown 
in Fig.\ref{fig:1}.  In both materials, there are three $t_{2g}$ bands 
right above the Fermi level, and for 
SrTiO$_3$, the band structure of the three band tight binding 
model is superposed to the 
original first principles band.
The band structure of the two materials 
look similar, but the band width is wider for KTaO$_3$ due to the 
widely spread nature of the $5d$ orbitals.

The calculated Seebeck coefficient 
for the two materials is shown in Fig.\ref{fig:2} 
against the temperature at $x=0.05$ and 
$x=0.1$ for SrTiO$_3$, and $x=0.009$ for KTaO$_3$.
We have chosen these $x$ to make comparison with the 
experiments\cite{Okuda_SrTiO3,Sakai_KTaO3}.
Here we take a rigid band approach, and 
assume that the hole concentration $n_h$ is 
equal to the La (SrTiO$_3$) and Ba (KTaO$_3$) content.

For SrTiO$_3$. The Seebeck coefficient at $300$K is 
$S(x=0.05) = -87\mu$V/K and $S(x=0.1) = -58{\rm \mu}$V/K. 
Experimentally, 
the Seebeck coefficient at 
$300$K is $S(x=0.05) = -147\mu$V/K and 
$S(x=0.1) = -88.7\mu$V/K\cite{Okuda_SrTiO3}. 
Thus the calculation result 
is somewhat reduced from the experimental result.
The reason for this is probably due to the reduction of the 
band width due to the strong correlation 
effect of the 3d orbitals.
In fact, it has been known from the comparison between band 
calculations and the angle resolved photoemission studies that the 
band width of the 3d electron materials is generally reduced by a factor of 
about two, and in fact taking this effect into account 
reproduces the experimental results of Na$_x$CoO$_2$ well\cite{Kuroki}.
If we calculate the  Seebeck coefficient at $300$K by reducing the 
band width by $50\%$ from the bare LDA result,    
we get 
$S(x=0.05) = -149\mu$V/K and $S(x=0.1) = -103\mu$V/K(Fig.\ref{fig:2}(a)),
which are in fact fairly close to the experimental values. 

As for KTaO$_3$,
the calculation of the Seebeck coefficient at $x=0.009$ gives 
$S(300$K$)=-160\mu$V/K.
This is roughly in agreement with the experimental result 
which is about $-200\mu$V/K\cite{Sakai_KTaO3}.
A reason why the bare LDA band structure gives good agreement with 
the experiments is because KTaO$_3$ is a $5d$ system, where the electron 
correlation effects are expected to be small compared to $3d$ systems 
like SrTiO$_3$.
In fact, for a number of rhodates, i.e. $4d$ systems, 
the Seebeck coefficient calculated from 
the bare LDA band structure gives fairly good agreement with 
the experiments\cite{Usui,AritaLiRh,Wilson}.

\section{Effect of the band multiplicity}
Having found that the experimentally observed 
Seebeck coefficient is roughly reproduced within the first principles 
band calculation + the Boltzmann's equation approach 
(with some additional consideration of band narrowing), 
we now explain why the 
Seebeck coefficient is large in these materials despite  
the relatively large conductivity. In other words, we seek for the 
origin of the large power factor $S^2\sigma$.

In the three orbital model, the Seebeck coefficient $S_{xx}$ is 
given as 
\begin{equation}
S_{xx} = \frac{1}{eT}\frac{K_{1}^{dxy} + K_{1}^{dyz} + K_{1}^{dzx}}{K_{0}^{dxy} + K_{0}^{dyz} + K_{0}^{dzx}}, 
\label{eq3}
\end{equation}
where $K_{n}^{dij}$ stands for $K_{n}$ of the $d_{ij}$ $(i,j=x,y,z)$ orbital. 
From eq.(\ref{eq2}), the group velocity $v^{dij}_{x}$ is the 
important factor in $K_n$. $v^{dxy}$ is equal to $v^{dzx}$ because 
$d(\varepsilon_{xy})/dx$ is equal to $d(\varepsilon_{zx})/dx$, 
so that $K_{n}^{dxy}=K_{n}^{dzx}$.
Also, $K_{n}^{dyz} \sim 0$ because $v^{dyz}$ is very small.
So the Seebeck coefficient is 
\begin{equation}
S_{xx} \sim \frac{1}{eT}\frac{2 K_{1}^{dxy}}{2 K_{0}^{dxy}} 
=  \frac{1}{eT}\frac{K_{1}^{dxy}}{K_{0}^{dxy}} = S^{dxy}_{xx}.
\label{eq4}
\end{equation}
Namely, the 
total Seebeck coefficient is equal to the Seebeck coefficient of the 
$d_{xy}$ single orbital system. 
On the other hand, the conductivity is 
\begin{equation}
\sigma =e^{2}(K_{0}^{dxy} + K_{0}^{dyz} + K_{0}^{dzx}) \sim 2e^{2}K_{0}^{dxy} = 
2\sigma^{dxy}.
\label{eq5}
\end{equation}
Therefore the power factor is
\begin{equation}
P_{xx} = \sigma S_{xx}^{2} \sim 2 \sigma^{dxy} (S^{dxy}_{xx})^{2} = 2P^{dxy}_{xx}.
\label{eq6}
\end{equation}
The left hand side here is the power factor of the three orbital system, 
while $P^{dxy}_{xx}$ in the right hand side is that of the $d_{xy}$ single orbital system.
Thus the multiplicity of the orbitals is advantageous for large power factor.
Note that the comparison here between the three orbital and the 
single orbital systems is given for the same number of electrons 
{\it per band}. If we present this relation between the three and one orbital 
systems using the {\it doping concentration} $x$, it should be given as 
$S_{xx}(3x)=S^{dxy}_{xx}(x)$ and $P_{xx}(3x) =2P^{dxy}_{xx}(x)$.
 
\begin{figure}
\includegraphics[width=8.5cm]{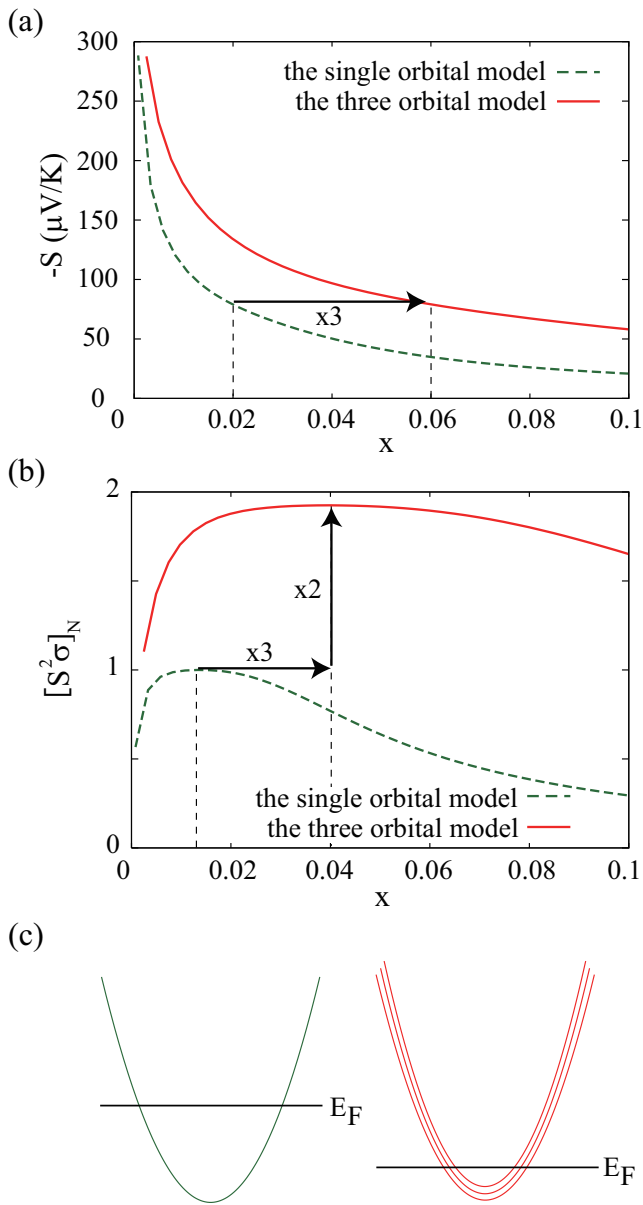}
\caption{\label{fig:3} (a) The Seebeck coefficient 
and  (b) the normalized power factor
of the single orbital model (dashed green) 
and the three orbital models (solid red) as functions of the doping rate $x$ at $300$K.
(c) Schematic figure of how the Fermi energy differs between single and multiband sysmtems.}
\end{figure}

In Fig.\ref{fig:3}, we show the actual calculation result of 
the Seebeck coefficient and the power factor 
(normalized at $x=0.13$ of the single orbital model)
of the $t_{2g}$ three orbital model of SrTiO$_3$  and
a single orbital model where only the $d_{xy}$ orbital is considered. 
It can be seen that the above relation is 
indeed satisfied. It is also worth noting that the doping dependence of the 
power factor is in striking agreement with the experimental 
observation (Fig.3 in ref.\cite{Okuda_SrTiO3}).
From this figure, we can see that for a fixed doping concentration, 
both the Seebeck coefficient and the power factor is 
larger for multiorbital systems than in single orbital ones.
This can intuitively be understood from Fig.\ref{fig:3}(c),
namely, the Fermi level tends to stay lower for systems with 
multiple bands for a fixed number of doped electrons, and lower 
Fermi level results in a large Seebeck coeffiecient, while the 
large number of electrons (due to the multiplicity of the 
bands) enhances the conductivity\cite{AritaLiRh}.
The present result  suggests that the band multiplicity 
is at least one of the main reasons why the Seebeck coefficient is 
large despite the metallic conductivity. The orbital degeneracy 
has been considered as a factor to obtain good thermoelectric 
properties in the context of entropy\cite{Koshibae,KoshibaePRL}, but 
we stress here that 
the present mechanism provides another way where the band multiplicity 
can play an important role\cite{AritaLiRh}.

\section{Effect of the band shape}

In the present materials, the density of states {\it per band} is 
not so large around the Fermi level.
This can roughly be understood in terms of the tight binding 
model. Namely, the tight binding model on a square lattice has 
electron hole symmetry when only the nearest neighbor hopping $t_1$ 
is considered. The introduction of the 
second nearest neighbor hopping $t_2$ breaks this electron-hole symmetry, 
and for $t_{2g}$ systems, this hopping integral usually has a 
negative sign when writing down the Hamiltonian in the form
$H=\sum_{ij} t_{ij} c_i^\dagger c _j$. 
When $t_2$ is negative,  
the density of states tends to be large in the upper half of the 
band and small in the lower half. In this sense, the effect of the 
so-called ``pudding 
mold type'' band\cite{Kuroki}, where a flat portion of the band has to be 
present near the Fermi level, is not relevant to the present 
electron doped materials. 
This can in fact be seen as follows. 
Since we have found that $S_{xx}\simeq S_{xx}^{dxy}$ (assuming same 
electron number per band) in the 
preceding section, we concentrate here on the $d_{xy}$ single orbital model
of SrTiO$_3$.
In this model, the nearest and second nearest neighbor hoppings 
of the MLWF tight-binding Hamiltonian are $t_{1}=-0.28$eV and 
$t_{2}=-0.078$eV.
To see how $t_2$ affects the Seebeck coefficient, 
we vary $t_{2}$ while fixing $t_{1}=-0.28$eV,  and 
calculate the Seebeck coefficient at $300$K as shown Fig.\ref{fig:4}(a).
It is found that the smaller $|t_{2}|$ is, 
the larger the Seebeck coefficient. This is because the lower part of the 
band (where the Fermi level exists) become less dispersive 
as $|t_2|$ is decreased 
when $t_2$ is negative. This can be seen in the calculation of the 
density of states (DOS) given in Fig.{\ref{fig:4}(b), namely,  
the DOS at the band bottom 
for $t_{2}=0$eV is about twice larger than for $t_{2}=-0.13$eV.
Thus the negative value of $t_2$ (i.e., the band shape) in 
SrTiO$_3$ is not faverable for thermopower, and the good 
thermoelectric properties seem to come mainly from the 
multiplicity of the bands.

We have also evaluated $t_1$ and $t_2$ for KTaO$_3$ from the obtained 
band structure as listed in table \ref{table1}  together with some related 
materials.
Although $t_1$ is much larger compared to that in 
SrTiO$_3$ as expected from the $5d$ nature, $t_2$ is not 
much changed, and 
the ratio $|t_2/t_1|$ is the smallest among the materials considered 
here. In fact, $|t_2/t_1|$ is also small in Zr and Nb compounds, namely 
4d systems with small number of electrons. So it seems that 
the ratio $|t_2/t_1|$ tends to be small for large principle 
quantum number. This trend can be considered as
another factor working favorable for the thermopower in 
KTaO$_3$ despite the wide band width.

\begin{figure}
\includegraphics[width=8.5cm]{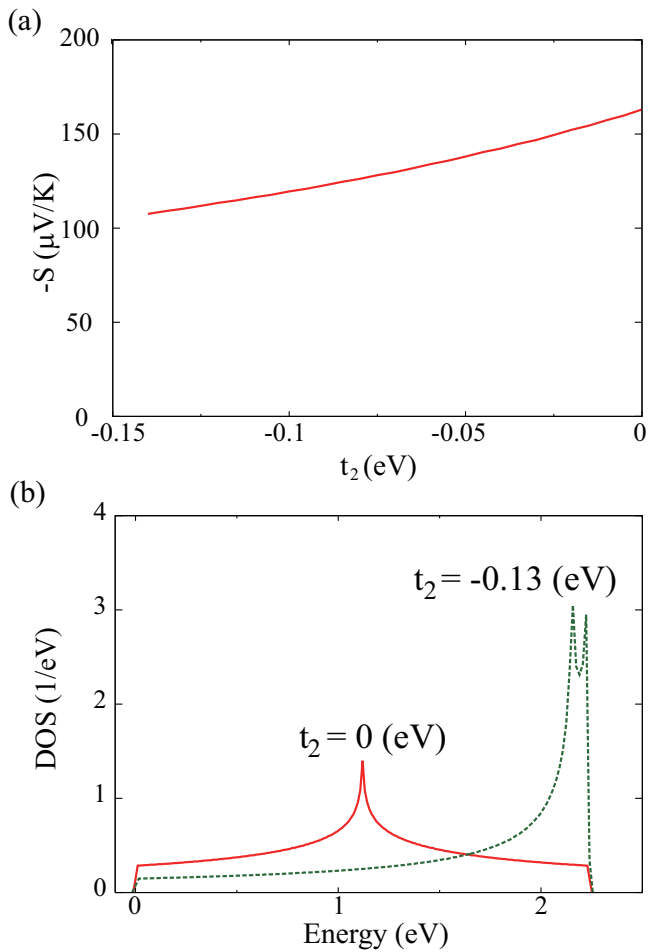}
\caption{\label{fig:4} (a) The Seebeck coefficient of the single orbital model for $x=0.017$, 
$t_1 = -0.28$eV and $T = 300$K plotted 
as functions of $t_{2}$. 
(b) The density of states at $t_1 = -0.28$eV, $t_2 = 0$eV (solid red) and 
$t_2 = -0.13$eV (dashed green).}
\end{figure}

\begin{table}
\caption{\label{table1}$t_1$, $t_2$ and $|t_2/t_1|$ 
obtained from constructing maximally localized Wannier orbitals.
In performing the band structure calculations, 
we have used the experimentally determined lattice parameters 
taken from the cited references.
For KTaO$_3$, we have obtained $t_{1}$ and $t_{2}$ by fitting 
the WIEN2k band structure with a tight binding model. 
}
\begin{ruledtabular}
\begin{tabular}{cccc}
 & $t_1$(eV) & $t_2$(eV) & $|t_2/t_1|$\\
\hline
PbTiO$_3$\cite{Glazer_PbTiO3} & -0.23 & -0.073 & 0.31\\
BaTiO$_3$\cite{King_BaTiO3} & -0.25 & -0.066 & 0.26\\
SrTiO$_3$\cite{Mitchell_SrTiO3_a} & -0.28 & -0.078 & 0.28\\
BaZrO$_3$\cite{Yamanaka_BaZrO3} & -0.40 & -0.081 & 0.20\\
NaNbO$_3$\cite{Shigemi_NaNbO3} & -0.45 & -0.091 & 0.20\\
KTaO$_3$\cite{Zhurova_KTaO3_a} & -0.52 & -0.094 & 0.18\\
BaMnO$_3$\cite{BaMnO3} & -0.17 & -0.067 & 0.41\\
\end{tabular}
\end{ruledtabular}
\end{table}


\section{Conclusion}
To conclude, we have studied the origin of the large Seebeck 
coefficient in SrTiO$_3$ and KTaO$_3$.
In SrTiO$_3$, from the first principles band calculation results, 
a tight-binding model is obtained via the maximally 
localized Wannier orbitals, and the Seebeck coefficient 
is calculated using the tight-binding model.
In KTaO$_3$, from the first principles band calculation results, 
the Seebeck coefficient is calculated using the BoltzTraP code.
In both materials, the large Seebeck 
coefficient despite the relatively large conductivity is 
largely due to the multiplicity of the bands, i.e., essentially 
the same value of the 
Seebeck coefficient is obtained for the same number of electrons 
{\it per band}, so that when the {\it total number of doped electrons} 
itself is the 
same, the Seebeck coefficient and thus the power factor are larger for 
multiple band systems. Also, we have examined the 
effect of the band shape. Although the negative $t_2$ value 
is not favorable for the electron doped thermoelectric materials, 
4d and 5d systems such as KTaO$_3$ tend to have similar $t_2$ values 
as in 3d systems despite the wide band width, 
and this can be 
another factor that is advantageous for good thermoelectric properties.

\begin{acknowledgments}
We acknowledge Yasujiro Taguchi for motivating us to start the present study.
We also thank Akihiro Sakai and Yoshinori Tokura for sending the 
preprint on the Ba-doped KTaO$_3$ prior to publication. 
The numerical calculations were in part performed at the 
Supercomputer Center, ISSP, University of Tokyo.
This work was supported by Grants-in-Aid from MEXT Japan.
H.U. Acknowledges support from JSPS.
\end{acknowledgments}


\end{document}